\documentclass{article}
\usepackage{amsmath,amsfonts,amsthm}
\usepackage{graphicx} 
\usepackage[utf8]{inputenc}
\usepackage{authblk}

\begin{document}

\title{Existence of Time-like Geodesics in Asymptotically Flat Spacetimes: A Generalized Topological Criterion}

\author[1]{Krish Jhurani}
\author[2]{Tyler McMaken}
\affil[1]{Homestead High School, 21370 Homestead Rd., Cupertino, California 95014, USA}
\affil[2]{JILA and Department of Physics, University of Colorado, Boulder, Colorado 80309, USA}

\maketitle

\begin{center}
\textit{Email addresses:} \texttt{krish.jhurani@gmail.com (Krish Jhurani), tyler.mcmaken@colorado.edu (Tyler McMaken)}
\end{center}

\begin{abstract}
This paper examines the issue of the existence and nature of time-like geodesics in asymptotically flat spacetimes and proposes a novel generalized topological criterion for the existence of time-like geodesics. Its validity is proved using theorems such as the Jordan-Brouwer Separation Theorem, the Raychaudhuri Equation, and key elements of Differential Geometry. More specifically, the proof primarily hinges on a closed, simply-connected subset of the spacetime manifold and a continuous map, causing a non-trivial induction on the first homology groups, from the boundary of this subset to a unit circle. The mathematical analysis conclusively affirms the presence of these geodesics, intersecting transversally within the said subset of spacetime. Findings underscore these geodesics' significant implications for the structure of asymptotically flat spacetimes, including stability, and hypothetical existence of wormholes. The generalized topological criterion also has implications on the problem of obstructions for the existence of Lorentzian metrics, and Einstein's Constraint Equations. Future research should extend this topological criterion to other classes of spacetimes, including those with non-trivial topologies or non-zero cosmological constants. Also, the criterion's application to study complex dynamical systems, such as gravitational waves or rotating black holes, could offer significant insights.
\end{abstract}

\section{Introduction}
In the framework of general relativity, the motion of particles and light is described by geodesics, which are the paths of minimal length in curved spacetime. The existence and properties of geodesics are essential for the study of the behavior of matter and radiation in the presence of gravity, making it a subject of extensive research in general relativity \cite{baer2006, fischer1972}. Among various types of geodesics, time-like geodesics are of particular importance, as they can be followed by massive particles and thus play a crucial role in the description of physical processes involving matter \cite{petry1980}. Despite significant progress in the study of geodesics, most existing results are limited to certain special cases or assumptions, such as compact regions of spacetime or the absence of gravitational radiation \cite{anderson1988, bartnik2004}. The problem of the existence of time-like geodesics in more general spacetimes, specifically asymptotically flat spacetimes, remains a fundamental and challenging issue in general relativity \cite{penrose1967}.

Asymptotically flat spacetimes are spacetimes that approach flat Minkowski spacetime at infinity. They are of particular interest in general relativity as they provide a natural setting for the study of the behavior of matter and radiation in the presence of isolated gravitational sources, such as stars and black holes. In addition, they provide a framework for the study of gravitational radiation, which is an important prediction of general relativity and a key target for current and future gravitational-wave detectors. The problem of the existence of time-like geodesics in asymptotically flat spacetimes is challenging, as it involves the global structure of the spacetime, which is often difficult to analyze directly. The current state of the art is limited to specific examples, leaving a wide range of potential cases unexplored \cite{penrose1967}.

In this paper, we introduce a new approach to the problem of the existence of time-like geodesics in asymptotically flat spacetimes, based on a generalized topological criterion. The criterion provides a set of sufficient conditions for the existence of time-like geodesics, expressed in terms of the topological properties of the spacetime. Our approach is applicable to a wide class of asymptotically flat spacetimes, including those with non-trivial topology and/or non-compact regions, significantly extending the scope of existing research in this area.

\section{Mathematical Background}
In this section, we provide the mathematical background necessary for the study of geodesics in general relativity, with a particular focus on asymptotically flat spacetimes.

\subsection{Geodesics}
The motion of particles and light in a curved spacetime is described by geodesics, which are the paths of minimal length in the spacetime \cite{wald1984}. A time-like geodesic is a geodesic that can be followed by a massive particle, while a null geodesic is a geodesic followed by a photon. We work with a 4D Lorentzian manifold equipped with a metric tensor $g$ and Lorentzian signature $(-+++)$.

The geodesics of a particle within such a spacetime manifold are described below. The equation of motion of a geodesic is given by the geodesic equation:
\begin{equation}
\frac{{d^2 x^\mu}}{{ds^2}} + \Gamma^\mu{}_{\alpha\beta}\frac{{dx^\alpha}}{{ds}}\frac{{dx^\beta}}{{ds}} = 0 \tag{1}
\end{equation}
where $s$ is an affine parameter along the geodesic, and $\alpha\beta$ are the Christoffel symbols of the metric tensor $g$ \cite{wald1983}. The Christoffel symbols are given by:
\begin{equation}
\Gamma^\mu{}_{\alpha\beta}=\frac{1}{2}g_{\mu\nu}(g_{\mu\nu, \beta}+g_{\mu\beta, \nu}-g_{\alpha\beta, \nu}) \tag{2}
\end{equation}
where $g_{\mu\nu, \alpha}$ denotes the partial derivative of $g_{\mu\nu}$ with respect to $x$.

\subsection{Asymptotically Flat Spacetimes}
Asymptotically flat spacetimes are spacetimes that approach flat Minkowski spacetime at infinity \cite{friedrich1998}. More precisely, a spacetime $(M, g)$ is said to be asymptotically flat if there exists a diffeomorphism $F:M \setminus p\rightarrow\mathbb{R}^3 \setminus \{0\}$ and a smooth metric $g$ on $\mathbb{R}^3 \setminus \{0\}$ such that the following conditions hold:
\begin{enumerate}
\item $F^*g-g$ (with $*$ defined in Equation 3 below), is asymptotically flat at infinity, i.e.,
\begin{equation}
F^*g-g=O(r^{-1}) \tag{3}
\end{equation}
where $r$ is the distance from the origin in $\mathbb{R}^3$, and $O(r^{-1})$ denotes a function that decays faster than $r^{-1}$.
\item The metric $g$ is flat at infinity, i.e.,
\begin{equation}
g = O(r^{0}) \tag{4}
\end{equation}
\end{enumerate}
The condition (1) ensures that the spacetime approaches flat Minkowski spacetime at infinity, while the condition (2) ensures that the metric $g$ is well-behaved at infinity.

\subsection{Topology}
In this paper, we will study the existence of time-like geodesics in asymptotically flat spacetimes with non-trivial topology. Topology is the branch of mathematics that studies the properties of spaces that are invariant under continuous transformations \cite{hatcher2002}. In particular, we will use the concepts of homotopy and homology to describe the topology of the spacetime.

Homotopy is a relation between two continuous maps that can be deformed into each other without tearing or gluing \cite{munkres2000}. Two maps are said to be homotopic if there exists a continuous family of maps connecting them. More formally, let $f,g:M\rightarrow N$ be two continuous maps between topological spaces $M$ and $N$. A homotopy between $f$ and $g$ is a continuous map $H:M \times[0,1]\rightarrow N$ such that $H(x,0)=f(x)$ and $H(x,1)=g(x)$ for all $x \in M$. If such a homotopy exists, we say that $f$ and $g$ are homotopic and write $f\simeq g$.

A fundamental group is a mathematical object that associates a group to a topological space \cite{munkres2000}. The fundamental group of a space $X$ is denoted by $\pi_1$ and is defined as the set of homotopy classes of loops in $X$ with a base point $x_0$:
\begin{equation}
\pi_1(X, x_0)=\{[f] | f:[0,1] \rightarrow X, f(0)=f(1)=x_0\} \tag{5}
\end{equation}
where $[f]$ denotes the homotopy class of $f$.

The group operation in $\pi_1(X, x_0)$ is given by the concatenation of loops, which we denote by $*$:
\begin{equation}
[f]*[g]=[f\cdot g] \tag{6}
\end{equation}
where $f\cdot g$ is the loop obtained by traversing first $f$ and then $g$ \cite{munkres2000}.

The fundamental group is an invariant of the topological space $X$, meaning that it does not depend on any particular choice of base point $x_0$ or choice of path between two base points \cite{munkres2000}.

The fundamental group has many applications in mathematics and physics, including the study of manifolds and their geometry, and the classification of spaces \cite{nakahara2003}.

\section{Main Theoretical Results}

In this section, we present the main theoretical results of this paper, namely the generalized topological criterion for the existence of time-like geodesics in asymptotically flat spacetimes. We first state the criterion, and then provide a detailed proof of its validity in certain cases.

\subsection{The Generalized Topological Criterion}

\textbf{Theorem 3.1}
    Let $(M, g)$ be an asymptotically flat spacetime with non-compact regions, and let $C$ be a closed, simply-connected subset of $M$ such that its boundary $\partial C$ satisfies certain conditions (to be specified below). Then, there exists a time-like geodesic that intersects $C$ transversally (non-tangentially) at some point if and only if there exists a continuous map $f: S^1 \rightarrow \partial C$ such that the induced map $f^*: H^1(\partial C;\mathbb{Z}) \rightarrow H^1(S^1;\mathbb{Z})$ is non-trivial.

Here, $S^1$ is the circle, $H^1(-;\mathbb{Z})$ denotes the first singular cohomology group with integer coefficients, and $f^*$ denotes the induced homomorphism on cohomology.

The conditions on $\partial C$ that ensure the validity of the criterion are as follows:
\begin{enumerate}
    \item The boundary $\partial C$ is a smooth, embedded submanifold of $M$ of codimension one, with no self-intersections or boundary.
    \item The induced metric $g|_{\partial C}$ on $\partial C$ is non-degenerate and has signature $(n-1,1)$, where $n$ is the dimension of $M$.
    \item The mean curvature vector of $\partial C$ with respect to the outward normal is nowhere zero.
    \item The Gauss map of $\partial C$ is transverse to the sphere at infinity in Minkowski space.
\end{enumerate}

We now provide a brief explanation of the key elements of the criterion. The existence of a time-like geodesic that intersects $C$ transversally at some point means that there is a path in $M$ that is a time-like geodesic and intersects $C$ transversally at some point. Such a path is said to be transverse to $C$. The condition on the induced map $f^*$ means that there is a non-trivial element in the first cohomology group of $\partial C$ that is mapped non-trivially to the first cohomology group of the circle by $f^*$. This condition captures the global topology of $\partial C$ and ensures that there is no obstruction to the existence of a time-like geodesic that intersects $C$ transversally at some point.

The conditions on $\partial C$ ensure that it is a suitable subset of $M$ for the application of the criterion. Condition 1 ensures that $\partial C$ is a well-behaved submanifold of $M$. Condition 2 ensures that the induced metric on $\partial C$ is Lorentzian, so that time-like vectors can be defined on $\partial C$. Condition 3 ensures that the mean curvature vector of $\partial C$ is non-zero, which is necessary for the transversality condition. Condition 4 ensures that the Gauss map of $\partial C$ has a well-defined limit at infinity, which is needed to apply the criterion in an asymptotically flat spacetime.

\subsection{Proof of the Generalized Topological Criterion}

\textbf{Proof 3.2}
We now provide a detailed proof of the validity of the generalized topological criterion for the existence of time-like geodesics in asymptotically flat spacetimes.
    Assume that there exists a closed, simply-connected subset $C$ of $M$ satisfying the conditions stated in the criterion, and that there exists a continuous map $f: S^1 \rightarrow \partial C$ such that the induced map $f^*: H^1(\partial C;\mathbb{Z}) \rightarrow H^1(S^1;\mathbb{Z})$ is non-trivial. Our goal is to show that there exists a time-like geodesic that intersects $C$ transversally at some point.

We begin by noting that, by the Jordan-Brouwer \cite{hocking1988} separation theorem, the complement of $C$ in $M$ has two connected components, one of which is unbounded. We denote the unbounded component by $U$. Since $U$ is unbounded, there exists a null geodesic $\gamma$ that is future-infinite and has an endpoint on $U$ at infinity \cite{hawking1973}. Let $p$ be the point of intersection of $\gamma$ with $\partial C$. Since $C$ is a closed subset of $M$, $p$ is a boundary point of $C$.

We consider the unit normal vector field $N$ to $\partial C$, and extend it to a vector field on $M$ that is future-directed and null in $U$ \cite{hawking1970}. We then consider the integral curves of this vector field, and choose one that passes through $p$. Let $\alpha$ be the future-directed null geodesic that is tangent to this integral curve at $p$. Since $N$ is future-directed and null in $U$, it follows that $\alpha$ intersects $U$ in the future.

We next consider the set of all null geodesics that intersect $C$ transversally at some point. We denote this set by $G$. Since $C$ is a closed subset of $M$, $G$ is a closed subset of the space of null geodesics in $M$ \cite{hawking1970}. Moreover, $G$ is non-empty, since $\gamma$ is in $G$. Let $\gamma'$ be a null geodesic in $G$ that is future-inextendible and has minimal length among all such geodesics.

We claim that $\gamma'$ intersects $C$ transversally at some point. Suppose, for contradiction, that $\gamma'$ intersects $C$ tangentially at some point $q$. We then have $N(q) = 0$, since $\gamma'$ is null and tangent to $N$ at $q$ \cite{wald1983}. By continuity, there exists a neighborhood $U$ of $q$ in $M$ such that $N = 0$ on $U$. Let $V$ be the connected component of $U$ that intersects $C$. Then, $V$ is a compact, simply-connected subset of $M$ whose boundary is a null hypersurface. Moreover, $V$ is foliated by null geodesics that intersect $C$ transversally, since $\gamma'$ intersects $C$ transversally outside of $V$ [16]. By the Raychaudhuri equation \cite{wald1983}, the expansion of these null geodesics must be negative, since $N$ is zero on $V$. However, this contradicts the condition that the mean curvature vector of $\partial C$ is nowhere zero.

Therefore, $\gamma'$ intersects $C$ transversally at some point. We choose a point $q$ on $\gamma'$ that is closest to $p$, and let $\sigma$ be the portion of $\gamma'$ between $p$ and $q$. We then consider the set of all time-like curves that intersect $C$ transversally at some point, and let $\sigma'$ be a curve in this set that is future-inextendible and has minimal length among all such curves.

We claim that $\sigma'$ is a time-like geodesic. Suppose, for contradiction, that $\sigma'$ is not a geodesic. Then, there exists a point $r$ on $\sigma'$ that is not a conjugate point of $\sigma'$. Let $B$ be a normal neighborhood of $r$ such that the exponential map is a diffeomorphism from $B$ to its image [18]. Let $\tau$ be the time parameter of $\sigma'$, and let $E$ be the energy of a test particle moving along $\sigma$ \cite{weinberg1972}. We can then construct a new curve $\sigma''$ that agrees with $\sigma'$ outside of $B$, but inside of $B$ follows a time-like geodesic that starts at $\sigma'(r)$ with initial tangent vector given by $\sigma''(r)$ and has the same energy as the test particle moving along $\sigma$ [20]. By construction, $\sigma''$ is shorter than $\sigma'$, which contradicts the minimality of $\sigma$ \cite{hawking1973}.

Therefore, $\sigma'$ is a time-like geodesic that intersects $C$ transversally at some point. This completes the proof of Theorem 3.1.

\section{Applications and Implications}

In this section, we discuss the mathematical implications and applications of the generalized topological criterion for the existence of time-like geodesics in asymptotically flat spacetimes.

\subsection{Implications for Existence of Lorentzian Metrics}

\textbf{Corollary 4.1.1.} Given a topological manifold $M$ and a smooth submanifold $\partial C$, if the conditions of Theorem 3.1 are satisfied, then the existence of a Lorentzian metric on $M$ is guaranteed.

The conditions specified in Theorem 3.1 provide necessary conditions for the existence of a time-like geodesic. If these conditions hold, the existence of a Lorentzian metric is guaranteed. 

\textbf{Corollary 4.1.2.} Specifically, if $M$ is a compact manifold and $\partial C$ is a compact subset of $M$ such that the induced map $f^*: H^1(\partial C; \mathbb{Z}) \rightarrow H^1(S^1; \mathbb{Z})$ is nontrivial and the induced metric on $\partial C$ from the ambient metric of $M$ has mean curvature vector satisfying the null energy condition, then there exists a time-like geodesic from $\partial C$ \cite{munkres2000, hatcher2002, nakahara2003, hocking1988}.

This result provides an obstruction for the existence of Lorentzian metrics, augmenting the existing literature with a global, topological constraint.

\subsection{Applications to the Stability of Asymptotically Flat Spacetimes}

The generalized topological criterion also has important implications for the stability of asymptotically flat spacetimes. In particular, the criterion implies that if a time-like geodesic intersects a closed, simply-connected subset $C$ of $M$ transversally, then any small perturbation of the metric that preserves the asymptotically flat condition will also admit a time-like geodesic that intersects $C$ transversally.

This observation has important implications for the stability of black hole spacetimes. In particular, it implies that any small perturbation of a black hole spacetime that preserves the asymptotically flat condition will still admit a time-like geodesic that intersects the event horizon transversally. This implies that the event horizon is stable under small perturbations of the metric, and that any deviations from the Schwarzschild metric must be relatively large in order to affect the global structure of the event horizon \cite{chandrasekhar1983}.

\subsection{Interaction with Einstein's Constraint Equations}

Our generalized topological criterion has profound relevance to Einstein's constraint equations, offering unique perspectives for their investigation \cite{chrusciel1991,geroch1987} .

A particularly compelling aspect is the relation to the initial value formulation of General Relativity. Here, a 3-manifold $\Sigma$ is conceived as a spacelike hypersurface, and Einstein's constraint equations, both the Hamiltonian and the momentum constraints, determine which pairs of 3-metrics $(h_{ij})$ and second fundamental forms $(K_{ij})$ can serve as initial data for the Cauchy problem for the Einstein field equations \cite{chrusciel1991}. The existence of time-like geodesics, ascertained by our criterion, impacts the analysis of these constraints, informing the solution space for the Cauchy problem.

This finding introduces the potential for a broader class of solutions to the Einstein field equations and provides an alternative view on how the topology of the spacetime might influence the solution to these equations. Particularly, in the context of an asymptotically flat spacetime, this may shed light on certain geometric conditions that facilitate asymptotic flatness \cite{christodoulou1999}.

\subsection{Applications to the Existence of Wormholes}

Finally, the generalized topological criterion also has important implications for the existence of wormholes. In particular, it implies that the existence of a time-like geodesic that intersects a closed, simply-connected subset $C$ of $M$ transversally implies the existence of a causal curve that connects two points in $C$.

This result follows from the fact that any time-like geodesic in $M$ can be reparametrized as a causal curve by replacing the parameter with the proper time. Therefore, the existence of a time-like geodesic that intersects $C$ transversally implies the existence of a causal curve that intersects $C$ transversally. Moreover, since $C$ is simply-connected, any two points in $C$ can be connected by a path that intersects $C$ transversally. Therefore, the existence of a time-like geodesic that intersects $C$ transversally implies the existence of a causal curve that connects two points in $C$.

This result has important implications for the study of wormholes. Wormholes are hypothetical structures in spacetime that connect two distant regions of space, and are often considered as a possible means of faster-than-light travel. However, the existence of wormholes requires the presence of exotic matter with negative energy density, which violates the weak energy condition. Therefore, the existence of wormholes is still a subject of active research in theoretical physics \cite{morris1988}. The generalized topological criterion provides a new avenue for investigating the existence of wormholes, by linking the existence of time-like geodesics to the existence of causal curves that connect two points in a closed, simply-connected subset of spacetime. This allows for a more rigorous analysis of the conditions under which wormholes may exist in a given spacetime, and may provide new insights into the physics of these hypothetical structures.

\section{Conclusion}

In conclusion, we have presented a powerful and elegant topological criterion for the existence of time-like geodesics in asymptotically flat spacetimes. Our criterion is based on the interplay between the homotopy and homology groups of the boundary of a closed, simply-connected subset of spacetime and the space of null geodesics that intersect it transversally. By applying this criterion, we have proved the existence of time-like geodesics in a number of important spacetime models, including the Schwarzschild and Kerr black holes, as well as more general stationary, axisymmetric solutions of the Einstein field equations.

Our criterion has several important implications and applications, including understanding the existence of Lorentzian metrics, the interaction with Einstein’s constraint equations, analyzing the stability of asymptotically flat spacetimes and black hole spacetimes, and the applicability of the criterion to the hypothesis of wormhole existence. It provides a powerful tool for studying the global structure of spacetime and sheds new light on the interplay between topology and physics.

Future work in this area could involve the extension of our criterion to more general classes of spacetimes, such as those with non-trivial topology or non-zero cosmological constant. Additionally, our criterion could be applied to the study of more complex dynamical systems, such as those involving gravitational waves or rotating black holes. Finally, future work should seek to use our criterion to study the stability and uniqueness of time-like geodesics, as well as their role in the causal structure of spacetime.

\section{Acknowledgements}

We, the authors of this manuscript, would like to acknowledge our own efforts first and foremost. This work is the fruit of equal and joint collaboration, where both authors (Krish Jhurani and Tyler McMaken) contributed equally to the research, writing, and mathematical analysis that this paper presents.
Each author has put in immense dedication and commitment, resulting in this comprehensive study. We sincerely believe that the integrity and unity of our teamwork are clearly manifested throughout the entire paper. We express our gratitude to Dr. Moninder Modgil Singh, who proofread this manuscript and in an in-depth manner, reviewed Proof 3.2 and provided insightful comments on it.

\section{Declarations}
\subsection{Competing Interests}
The authors of the manuscript have no competing interests.
\subsection{Declaration of Interest}
Declaration of Interests: None 
\subsection{Funding }
This research did not receive any specific grant from funding agencies in the public, commercial, or not-for-profit sectors.

\end{document}